%% file: main.tex
\documentclass[runningheads]{llncs}
\usepackage[T1]{fontenc}
\usepackage{graphicx}
\usepackage{xcolor}
\usepackage[normalem]{ulem}
\usepackage{hyperref}
\usepackage[font=small,labelfont=bf,labelsep=colon]{caption}
\usepackage{subcaption}
\begin{document}
\title{A Template-Driven Platform for Contextualised Researcher Profiles}


%
%
\author{
Serafeim Chatzopoulos\inst{1}\orcidID{0000-0003-1714-5225}
\and
Paris Koloveas\inst{1,2}\orcidID{0000-0003-2376-089X}
\and
Kleanthis Vichos\inst{1}\orcidID{0000-0002-8955-9489}
\and
Dionysis Diamantis\inst{1}\orcidID{0009-0002-3272-1294}
\and
Thanasis Vergoulis\inst{1}\orcidID{0000-0003-0555-4128}
}
\authorrunning{S. Chatzopoulos et al.}
%
\institute{
IMSI, Athena RC, Athens, Greece\\
\email{\{schatz, pkoloveas, kvichos, dionysis.diamantis, vergoulis\}@athenarc.gr}
\and
University of the Peloponnese, Tripolis, Greece
}
\maketitle              
\begin{abstract}
Modern researchers engage in diverse activities, assume multiple contribution roles, and produce a variety of outputs beyond traditional publications. This broader view of research contributions is increasingly recognised by responsible research assessment initiatives. However, existing researcher profiling platforms remain largely focused on publications and publication-centric indicators, offering limited support for contextualised and multi-dimensional representations of research careers. This paper presents BIP! Scholar, a platform that supports flexible researcher profiling through a template-driven approach. Researchers can create profiles tailored to different presentation or assessment contexts using track-based, narrative-style, or hybrid templates which support the representation of diverse outputs, contribution roles, and broader research activities. The platform also supports research assessment experts who wish to design and evaluate experimental profile templates. 
\end{abstract}
\input{sections/1_intro}
\input{sections/2_features}

\input{sections/3_architecture}
\input{sections/4_demo}
\input{sections/5_related}

\input{sections/6_conclusion}

\subsection*{Acknowledgments}

{\footnotesize
This project has received funding from the European Union’s Horizon 2020 research and innovation programme under grant agreement No 101095129 and 101129744.

\vspace{0.3em}
\centering
\includegraphics[width=0.08\linewidth]{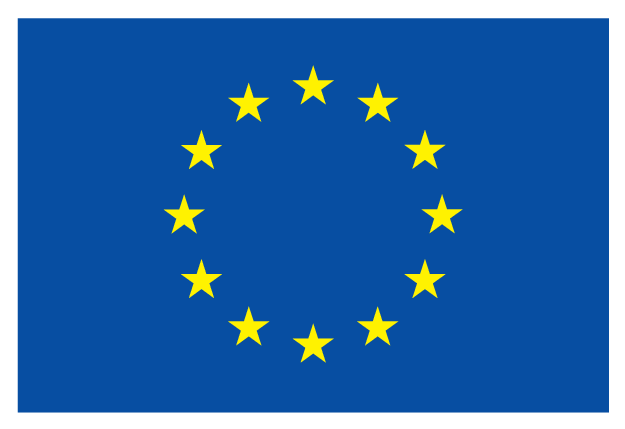}
\par
}

\bibliographystyle{splncs04}
\bibliography{thebib}

\end{document}

%% file: sections/1_intro.tex
\section{Introduction}
\label{sec:intro}

For decades, publications were regarded as the primary form of research contribution, while authorship served as the main mechanism for crediting research work. However, over the last fifty years, the research landscape undergone a profound transformation: multi-authorship and interdisciplinary collaboration have become increasingly prevalent~\cite{brand2015credit}, while it became common for modern researchers to engage in a much broader spectrum of activities~\cite{HK_principles,hidden_ref,Farrell2026-zu}, extending beyond theorising, experimentation, and manuscript writing (all strictly tied to publications) to include various other types of activities such as data curation, software development, mentoring, and the dissemination of research outcomes to both the scientific community and the society at large.
This broader understanding of research contributions is increasingly recognised by contemporary initiatives for responsible research assessment, such as DORA~\cite{dora2012} and CoARA~\cite{coara2022}.

However, widely adopted researcher profiling platforms, such as Google Scho\-lar and ResearchGate, primarily emphasise publications and citation-based indicators (e.g., the h-index) as the canonical representation of a researcher’s career. While other platforms, such as ORCID~\cite{haak2012orcid}, extend coverage to additional research activities (e.g., funded projects, peer reviews, invited talks, etc.), they still rely on largely static and uniform profile structures that are applied across disciplines and use cases. As a result, current researcher profiling platforms provide only limited support for the contextualised and multi-dimensional representation of research careers, failing to adequately reflect differences across disciplines, career stages, institutional contexts, and individual contribution roles, as advocated by responsible research assessment initiatives.

This limitation has become increasingly significant as contemporary research practices continue to evolve. Modern research is frequently conducted within heterogeneous and interdisciplinary teams that include a broad range of contributors, such as scientific technicians, data stewards, and research software engineers, whose work is often insufficiently represented by conventional profiling models~\cite{hidden_ref}. At the same time, researchers are required to present different facets of their work depending on context: indicatively, funding applications may emphasise societal impact, while collaboration and promotion cases may require evidence of leadership and mentoring. Consequently, a fundamental challenge persists: there is no one-size-fits-all representation of a research career in an era where scholarly contributions are increasingly diverse and extend far beyond traditional publications. Despite this growing need for flexibility and contextualisation, there is currently no widely adopted platform that supports the systematic design, evaluation, and deployment of multiple customisable profile templates aligned with evolving research assessment practices.

This paper presents BIP! Scholar,\footnote{BIP! Scholar: \url{https://bip.athenarc.gr/scholar}} an open-source platform\footnote{Parent code repo: \url{https://github.com/athenarc/bip-services}} designed to support richer and more flexible representations of researcher contributions through a template-driven profiling approach. 
The platform enables researchers to create and share profiles tailored to different presentation and assessment contexts using a variety of profile templates. Profile content is generated through a combination of automatically retrieved metadata from open scholarly sources and researcher-provided input, allowing users to highlight diverse outputs (e.g., datasets, software) and broader research activities. 
In addition, the platform supports research assessment experts who wish to experiment with and evaluate alternative profiling approaches. These users can design and deploy custom profile templates, share them with volunteer researchers, and collect feedback.



%% file: sections/2_features.tex
\section{Platform Overview and Key Features}
\label{sec:features}

BIP! Scholar is a researcher profiling platform designed to support and inclusive, transparent, and fair representation of research careers through a variety of profile templates. The platform is implemented as open-source software and builds upon open scholarly knowledge graphs, such as the OpenAIRE Graph~\cite{manghi2022openaire}. Its key features are elaborated in the following sections. 

\subsection{Creation, management, and sharing of researcher profiles}
\label{sec:prof}

\begin{figure}[t]
  \centering
  \fbox{\includegraphics[width=0.99\textwidth]{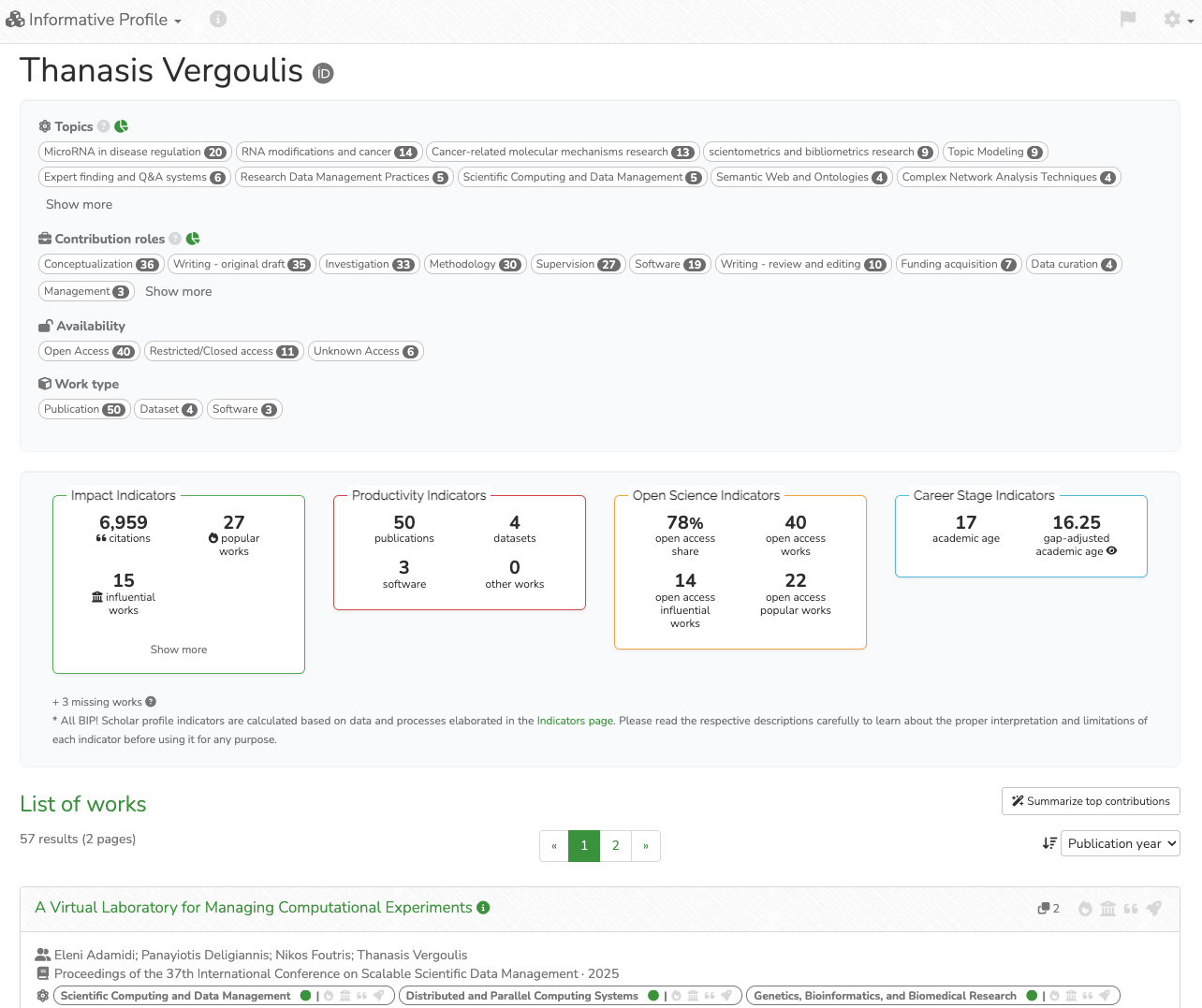}}
  \caption{Example profile based on the ``Informative Profile'' template of BIP! Scholar.}
  \label{fig:profile-first}
\end{figure}

BIP! Scholar enables researchers to create, manage, and share profiles tailored to different presentation and assessment contexts through a diverse set of profile templates. Researchers may choose to keep their profiles private for self-monitoring and personal reflection purposes, or make them publicly accessible to facilitate discovery and external visibility.
The templates provided by the platform vary substantially in both structure and emphasis. Some prioritise structured representations of scholarly output through detailed lists of contributions (e.g., the ``Informative Profile'' template, see also Figure~\ref{fig:profile-first}), others adopt a predominantly narrative-oriented format (e.g., the ``Résumé for Researchers'' template), while hybrid approaches combine narrative descriptions with structured contribution lists (e.g., the ``Brief Research CV'' template).

Many templates incorporate dynamic and interactive elements, enabling viewers to explore and refine profile contents according to specific criteria, such as the research topic or access license. This allows profiles to adapt to different exploration and assessment needs.
For example, Figure~\ref{fig:profile-first} presents a profile generated using the ``Informative Profile'' template. The profile includes an automatically generated list of scholarly contributions (publications, datasets, software tools, and other research outputs) together with researcher-level indicators and interactive filtering facets. When a viewer selects a specific selection of filters (e.g., a combination of topics), both the displayed contribution list and the associated indicators are dynamically updated to reflect only the filtered subset of works.

To reduce the manual data-entry burden, profile authoring is supported by two complementary semi-automated content-generation mechanisms. Contribution lists are populated through aggregation from open scholarly sources (see Section~\ref{sec:metadata-proc}), while an AI-assisted component produces, where applicable, candidate prose summaries that support the drafting of narrative sections. 

\subsection{Discovery and exploration of researcher profiles}
\label{sec:pdiscovery}

BIP! Scholar provides a search functionality for discovering publicly available researcher profiles. The search interface is accessible from the top section of the platform’s main page, allowing users to search for researchers by entering a full or partial name query. 
Upon submitting the query, the platform retrieves matching researcher records associated with public profiles. Users can then browse the returned results and open the profile of the researcher of interest, when available.

\begin{figure}[t]
  \centering
  \fbox{\includegraphics[width=0.99\textwidth]{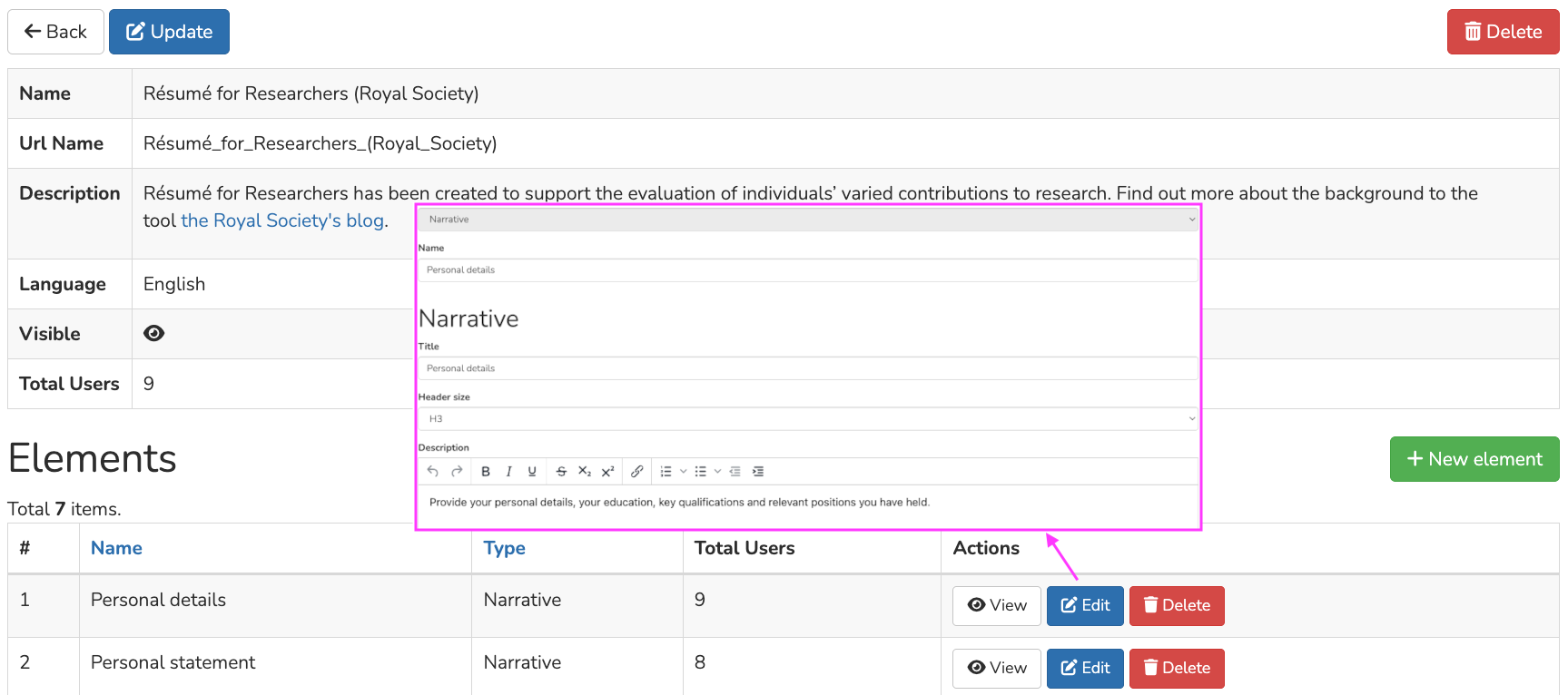}}
  \caption{BIP! Scholar template editor interface.}
  \label{fig:template-creator-interface}
\end{figure}

\subsection{Design, customisation, and evaluation of researcher profile templates}
\label{sec:tedit}

The platform also provides a template editor (Figure~\ref{fig:template-creator-interface}) through which assessment experts can compose new profile templates from a variety of element types, including narrative sections, indicator collections, contribution lists, dropdown selectors, and other structured fields. The resulting templates can be piloted privately with volunteer researchers and iteratively refined based on real-world usage and feedback.
To inform this refinement process, template creators are further supported by monitoring information, including completion statistics and coverage across template elements (see ``Total Users'' column of Figure~\ref{fig:template-creator-interface}), as well as structured feedback collected from participating researchers. Together, these signals provide the empirical basis on which template editors evaluate the effectiveness of their templates and guide subsequent revisions.
Mature and validated templates can subsequently be incorporated into the platform’s default template collection, making them available to the wider research community.

%% file: sections/3_architecture.tex
\section{System Architecture}
\label{sec:system}

\begin{figure}[t]
  \centering
  \includegraphics[width=\textwidth]{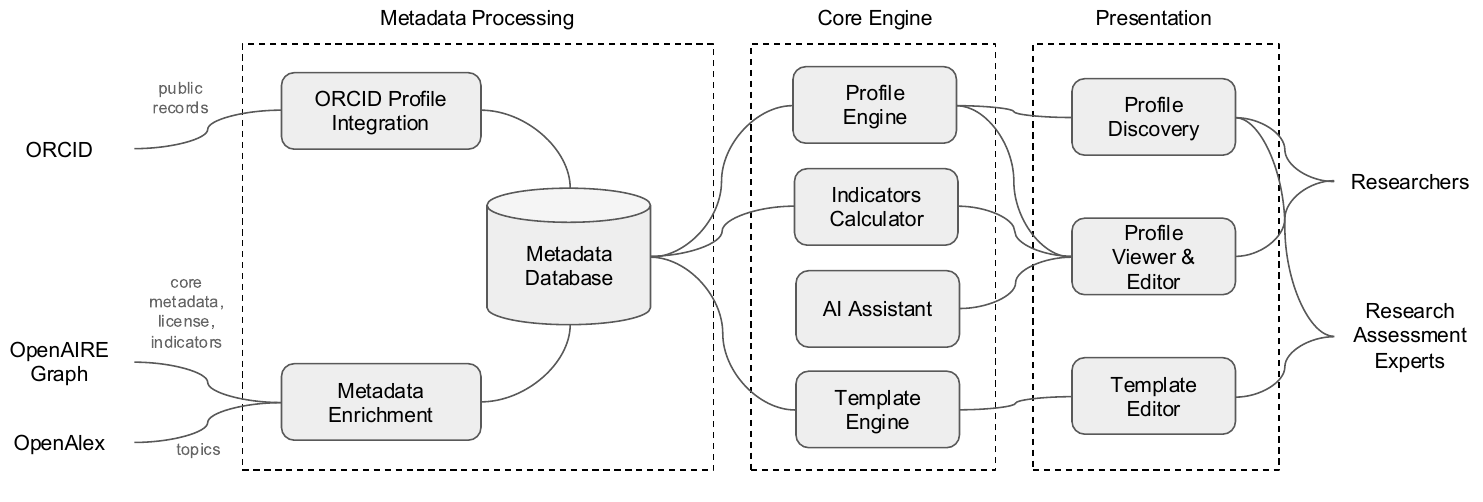}
  \caption{Overview of the BIP! Scholar architecture.}
  \label{fig:arch}
\end{figure}

The architecture of BIP! Scholar is structured into three main layers: a metadata processing layer, a core engine layer, and a presentation layer. Figure~\ref{fig:arch} presents the overall architecture and the flow of data between them. In the following sections each architecture layer is further elaborated.

\subsection{Metadata Processing}
\label{sec:metadata-proc}

This layer is responsible for collecting, storing, and enriching the research-profile metadata underlying the BIP! Scholar platform. Its \emph{ORCID Profile Integration} component imports public records of research outputs (publications, datasets, etc) from researchers who choose to link their ORCID accounts and create a profile within BIP! Scholar. The retrieved records are further enriched by the \emph{Metadata Enrichment} component using metadata harvested by the OpenAIRE Graph~\cite{manghi2022openaire} and OpenAlex~\cite{priem2022openalex}, two of the largest open scholarly knowledge graphs currently available. Most types of metadata (e.g., author lists, titles, venues, citation-based indicators, licesing information) are collected by the OpenAIRE Graph, while research topics are derived from OpenAlex.\footnote{Level-2 topic classifications are used for enrichment.}
Finally, the \emph{Metadata Database}, implemented as a relational database, serves as the central repository for storing and managing the collected and enriched metadata.

\subsection{Core Engine}
\label{sec:core-eng}

This layer implements the main operational logic of the platform. The \emph{Profile Engine} is responsible for managing  structured researcher profiles, ensuring that the user actions performed through the \emph{Profile Viewer \& Editor} and the \emph{Profile Discovery} components (Section~\ref{sec:presentation}) are executed successfully and, where necessary, properly reflected in the \emph{Metadata Database}. 

Similarly, the \emph{Template Engine} supports the creation, configuration, and maintenance of researcher profile templates by implementing the actions initiated through the \emph{Template Editor} (Section~\ref{sec:presentation}). This also involves handling functionalities related to template validation, including feedback collection and usage analytics generation to support template evaluation and refinement.

The \emph{Indicators Calculator} component is responsible for computing a variety of researcher-level indicators including output counts (e.g., number of publications), aggregated impact indicators (e.g., based on the popularity~\cite{vergoulis2021bip}, influence~\cite{vergoulis2021bip}, and citation count of the researcher's outputs), open science practicing indicators (including open access share across outputs), and career stage 
indicators (e.g., academic age). 

Finally, the \emph{AI Assistant} component implements all AI-supported functionalities of the BIP! Scholar platform. When the corresponding option is enabled by the user, the component can generate AI-assisted summaries of contribution lists upon request, thereby reducing the effort required to bootstrap narrative descriptions of a researcher’s work and achievements. These generated narratives can serve as a starting point or source of inspiration for researchers when drafting descriptions of their research activities and 
contributions, which may be required by certain profile templates. The feature 
relies on a configurable large language model (currently DeepSeek V4 Flash), 
and uses carefully engineered prompts\footnote{Prompts: \url{https://github.com/athenarc/scientific-summarization-api/blob/main/system_prompts.yaml}.} optimised for performance.

\subsection{Presentation Layer}
\label{sec:presentation}

This layer provides the user interface of the BIP! Scholar service, addressing distinct requirements of its end users, while orchestrating the interactions with the core engine components and, indirectly, the metadata processing layer. 
More specifically, the \emph{Profile Viewer \& Editor} component allows researchers to create and manage their profiles using various templates available in the platform's default collection. In addition, it allows users to view public researcher profiles that have been shared with them. 
The \emph{Template Editor} component supports research assessment experts in designing, configuring, and maintaining researcher profile templates. Moreover, the interface facilitates the collection of targeted feedback from volunteer researchers participating in the validation process and provides analytics (e.g., template usage and completion statistics) that help template creators monitor the progress and effectiveness of the validation activities. These insights support the evaluation and iterative refinement of the templates. 
Finally, the \emph{Profile Discovery} component enables the platform end users to search for and explore publicly available profiles created within the platform. 

%% file: sections/4_demo.tex
\section{Demonstration Scenarios}
\label{sec:demo}

During the demonstration session, the audience will have the opportunity to interact directly with the BIP! Scholar platform and explore its key features. While participants will be free to navigate and use the platform as they wish, the demonstration will be structured around two main scenarios, each showcasing a core functionality of the platform.\footnote{For the anonymous review process, we provide a demonstration account (user: `reviewer' / password: `reviewer') for testing the provided scenarios anonymously.}

In the \emph{first scenario}, participants will use the \emph{Profile Discovery} functionality to search for public researcher profiles, using as an example one of the presenters' fully completed profiles. They will explore these profiles through the \emph{Profile Viewer}, examining how the same activities and contributions can be presented using different templates. The presenter will then demonstrate how targeted enrichments, such as contribution roles for publications or narrative content in specific sections, can be added through the \emph{Profile Editor}, providing participants with an overview of the platform’s profile customization capabilities.

In the \emph{second scenario}, participants will adopt the role of a research assessment expert experimenting with a new researcher profile template. Using the template editor, they will create a simple template by combining various available element types (see also Section~\ref{sec:tedit}). The template will then be deployed as a draft, enabling participants to examine it from the researcher’s perspective, review usage statistics and feedback, and iteratively refine it.

%% file: sections/5_related.tex
\section{Related Work}
\label{sec:related}

Most existing \emph{researcher profiling platforms} are services provided by organisations focused on aggregating scholarly metadata. The resulting profiles primarily consist of publication lists enriched with basic metadata (e.g., research topics) and indicators derived from citation network analysis. These platforms also aggregate publication-level indicators to compute researcher-level indicators, such as the h-index, often accompanied by simple visualizations. Representative examples include Google Scholar Profiles, Scopus Author Profiles, and Web of Science Researcher Profiles. Similar services are also offered by academic social networks such as ResearchGate and Academia.edu.

However, this strong emphasis on publications and citation metrics fails to capture the diversity of activities and contributions that characterize contemporary research practice. These limitations have been increasingly recognized by initiatives advocating for responsible research assessment, such as DORA~\cite{dora2012} and CoARA~\cite{coara2022} (see also Section~\ref{sec:intro}).
ORCID~\cite{haak2012orcid}, which has emerged as the dominant persistent identifier infrastructure for researchers, partially addresses this need by aggregating a broader range of scholarly activities beyond publications. ORCID records can also provide contextual information, such as contributor roles. Nevertheless, the ORCID profile model remains structurally flat and largely context-agnostic. It does not support alternative profile templates tailored to different career paths, assessment purposes, or presentation perspectives. Instead, ORCID enforces a fixed profile structure. 
More recently, OpenAIRE introduced a pilot proof-of-concept profiling platform called MyResearchFolio,\footnote{MyResearchFolio: \url{https://researcherprofile.openaire.eu/}} which aims to integrate non-traditional research outputs into richer researcher narratives. However, the platform is currently limited to a small number of demonstration profiles and does not yet provide an operational environment for user registration or profile management. Furthermore, similarly to ORCID, it supports only a single-profile representation and does not enable multiple configurable templates adapted to 
different assessment or communication contexts.

In this landscape, BIP! Scholar already occupies a distinct position by building upon an earlier researcher profiling service that enabled the creation of structured researcher profiles covering diverse scholarly activities and contributions~\cite{vergoulis2022bip}. 
The present work extends this foundation by introducing support for multiple profile templates, enabling different contextualised representations of the same researcher for distinct assessment and communication purposes. This evolution transforms the system from a single-profile model into a highly configurable profiling framework that supports the design, deployment, and refinement of templates tailored to diverse research assessment needs. 
Importantly, BIP! Scholar is openly accessible for registration and released as open-source software developed by a non-profit organisation, in contrast to many established profiling platforms operated by commercial providers.

%% file: sections/6_conclusion.tex
\section{Conclusion}
\label{sec:conclusion}

This paper presented BIP! Scholar, an open-source platform for the flexible, template-driven representation of research contributions. The system enables researchers to maintain profiles based on multiple templates, each tailored to specific presentation or assessment contexts, while also providing research assessment experts with the infrastructure needed to design, pilot, test, and iteratively refine experimental researcher profile templates.